# Spike sorting using non-volatile metal-oxide memristors

Isha Gupta[a,*], Alexantrou Serb[a], Ali Khiat[a], Maria Trapatseli[a], Themistoklis Prodromakis[a].

[a]Department of Electronics and Computer Science, Faculty of Physical Science and Engineering, University of Southampton, University Road, SO17 1BJ, Southampton, United Kingdom. Corresponding Author[*]: Isha Gupta (Email: I.Gupta@soton.ac.uk)

**Electrophysiological techniques have improved substantially over the past years to the point that neuroprosthetics applications[1] are becoming viable[2,3] . This evolution has been fuelled by the advancement of implantable microelectrode technologies[4,5] that have followed their own version of Moore's scaling law[4,6]. Similarly to electronics, however, excessive data-rates[7] and strained power budgets require the development of more efficient computation paradigms for handling neural data in-situ; in particular the computationally heavy task of events classification. Here, we demonstrate how the intrinsic analogue programmability of memristive devices[8–10] can be exploited to perform spike-sorting[11]. We then show how combining memristors with standard logic enables efficient in-silico template matching. Leveraging the physical properties of nanoscale[12] memristors allows us to implement ultra-compact analogue circuits for neural signal processing at the power cost of digital.**

Spike sorting is the procedure of identifying the activity of individual neurons from data collected through electrophysiological experiments[13,14,15,16]. Typically this involves processing raw neuronal data by first detecting the presence of action potential (spiking) activity, then extracting appropriately chosen features and finally, clustering the results; each cluster corresponding to an individual neuron. Memristive devices can inherently act as thresholded integrators[17]. When presented with an input voltage waveform the devices accumulate changes in resistive state linked to the instantaneous signal magnitude and polarity, so long as this exceeds the device threshold. We recently exploited this property for detecting neuronal spiking activity[8] while filtering out background noise.

The combination of any programming signal amplitude and polarity that induces analogous resistive switching can be treated as a spike feature, thus rendering spike sorting at the memristor level possible. A simplified diagram of a single memristor-based spike sorter channel is shown in Fig. 1(a). Analogue input neural data is subjected to suitable amplification and is then fed into the memristive device (see methods and Supplementary Figure 1). The amplification step ensures that action potential contributions to the input waveform exceed the device's thresholds, whilst background noise does not. Periodically the memristor's resistive state is assessed and the differences between consecutive readings are recorded (Supplementary Figure 1(b)). This is illustrated in Fig. 1(b-g) and Supplementary Figure 2, where the prominent peaks in the three example neural signals cause distinctive changes in the resistive state of the device under test. In order to avoid accumulating resistive state changes to the point of saturation[9], the devices are regularly reset to a suitable resistive state baseline[9] (also see Supplementary Figures 3).

The capability of single memristive devices to act as spike sorting elements was experimentally demonstrated using a commercially available memristor characterisation instrumentation (see methods section and Supplementary Figure 4). Publicly available simulated neural signals constructed based on electrophysiological recordings from the cortex and basal ganglia[15] were used as the input data for the memristive spike sorter (Supplementary Figure 5). The data contained three distinct single-unit activity waveform prototypes overlaid on a noise background (Supplementary Table 1,2). The instrument applied suitable amplification and relayed these waveforms to stand-alone devices.

For the first experiment, averages of each of the three single-unit waveforms were obtained by pooling ten random instances from each class. These were then arranged in a spike triplet (Fig. 2(a)) and Supplementary Figure 6, at the end of which a reset pulse was appended. Ten spike triplets were sequentially fed into the test memristor and its resistive state was regularly assessed 3 times during each triplet, producing results similar to Fig. 1(b-g) (amplification gain: -1.3 offset: -0.63V). Plotting the change in resistive state between every pair of consecutive measurements vs. the resistive state in the first measurement of the same pair results in Fig. 2(b). These results capture the impact of each spike's strength on setting the memory state of the device with respect to its initial state, as exemplified by the shaded spikes in Fig. 2(a) and their corresponding R(ΔR) points in Fig. 2(b) (for raw data example see Supplementary Table 3). The clear clustering of data-points associated with the three distinct spike waveform prototypes demonstrates that the memristive device is capable of intrinsically performing spike sorting consistently. As the waveform input to the device is identical for each triplet, any variation in R(ΔR) response arises mainly from device variability.

For the second experiment, the triplets were constructed from individual spike instances (not averaged prototypes) and their order was randomised as shown in Fig. 2(c) and Supplementary Figure 7. This setup accounts for background noise-induced spike shape variability[15]. Results are shown in Fig. 2(d) and Supplementary Table 4, where the intrinsic variability in device behaviour is compounded by the variability in the spike waveforms. Notably, despite the fact that clustering is no longer as clear as in Fig. 2(b), it is still possible to linearly separate the majority of events. The misclassified spike that caused strong change in resistive state (enclosed in a box) was the result of the event instance containing two spikes in close succession (main spike plus a stray spike).

One of the most popular approaches to performing spike sorting is the template matching technique[18]. It is a very powerful yet computationally intensive technique that involves regularly sampling and digitising data arriving from the electrophysiological set-up and comparing small snippets of consecutive samples (typically 10-20[19]) against a set of stored templates. Its strength stems from the fact that whenever a match is found the system registers the occurrence of a spike and the matching template ID, thus simultaneously providing spike timing and identification information. Here, we implement the template matching technique with memristive technologies directly in the analogue domain, using the devices programmability for storing the templates close to the site of computation. This obviates both the requirement for an analogue-to-digital converter (ADC) at the high bandwidth input signal line and the need for a dedicated memory circuit block for storing the templates.

This concept is realised via the circuit topology shown in Fig. 3(a). We refer to this circuit as a 'texel' (template pixel) that consists of a memristor and six transistors and its output current $I_{out}$ depends on the proximity of the input voltage value $V_{IN}$ to a stored value that is determined by the resistive state of memristor R1. When $V_{IN}$ is close to stored value $V_{pk}$, $I_{out}$ increases, therefore the texel acts as an input-vs-stored voltage distance-calculating circuit. Internally it consists of an enhanced inverter followed by an output stage. The output stage forces $I_{out}$ to peak whenever internal node voltage $V_{MID}$ reaches some optimal value $V_{OPT}$ determined by the transistor sizings in the output stage, but the value of $V_{IN}$ that forces $V_{MID}$ to that optimal level (a.k.a. the stored value $V_{pk}$) is determined by the memristor R1. This process is shown in Fig. 3(b) by illustrating measured transfer characteristics of a discrete texel circuit.

The texel concept was evaluated by assembling an array of four texels, feeding them with nine neural spike waveforms from the same database[15] (Supplementary Figure 5(a)) and summing the current outputs of each texel down a common load resistor, as shown in Fig. 3(e,f). Three spike instances were chosen from each class presented in Fig. 1: a low (L), a medium (M) and a high (H) instance corresponding to spikes exhibiting lower than, similar to or higher than class-average voltage levels (see methods and Supplementary Figure 8). The voltage level at the system's $V_{OUT}$ terminal is linked to the degree of matching between the input vector **k** and the stored template and was directly used as a matching degree metric. Due to the similarity between the H instance of class 1 and the L instance of class 2 and the limited resolution of our instrumentation, the experiment for these two instances was ran only once with a common input vector **k** (see Supplementary Table 5). Results are moreover shown in Fig. 3(g) for a texel array set up to discriminate for class 2 spikes. Even using only four samples from each waveform (marked in Fig. 3(e)) strong discrimination between templates is clearly achieved.

In conclusion, we demonstrate that memristor-based spike sorting systems are promising candidates for future brain-machine interfaces. The spike sorting ability of the concept systems presented here is compounded by the positive downscaling prospects of memristive technologies, both in area[12] and power[20]. Our results prove that single nanometre scale devices can capture enough input signal information for encoding distinct spike classes at no extra power (~100nW)[21] or area cost to what we have shown previously for spike detection[17,21]. Moreover, the proposed memristor-based texel architecture enables carrying out the computationally intensive template matching-based spike sorting[19] on a low component count, low parasitic capacitance system that essentially consists of a few, modified inverters and operates in the analogue domain on a digital power budget (see Supplementary Figures 9,10,11 and Supplementary material 1). Our results bring new application prospects for memristive devices, diversifying from conventional digital memory applications towards enabling active neural interfacing technologies that are very much needed for realising the electroceuticals vision[22].

**Methods**

**Memristive devices:** For all experiments in this work bilayer metal-oxide memristors were used with stack configuration Pt/AlOx/TiOx/Pt (10/4/40/10 nm). The devices featured an area of 20x20 um$^2$. Patterning was carried out using conventional optical lithography and lift-off processes. The electrodes were deposited using e-beam evaporation while the oxides were fabricated using magnetron sputtering. Electrical characterisation was carried out using a memristor characterisation and testing platform featuring ArC instruments Ltd. Technology (http://www.arc-instruments.co.uk/). Before use all devices had to be electroformed; a one-time process that electrochemically activates the memristor[23]. Our test devices typically electroform once at 8-12V. Once electroformed, the devices show reliable, well-behaved switching in the [+1.2, +2.5]V and [-1.2, -2.5]V ranges. For this work the devices were used in the 5-20 kΩ resistive state operating range.

**Neural data source:** The neural data used for our experiments is publically available from the univ. of Leicester, R. Q. Quiroga group (http://www2.le.ac.uk/centres/csn/software)[15]; dataset no. 2. It consists of a simulated neural recording synthesised using three distinct singe-unit activity templates (extracted from measured data) overlaid on top of background noise. We refer to each unique combination of a standardised spike waveform plus noise as a 'spike instance'.

**Memristive spike sorter signal processing:** Input neural recording data was processed as illustrated in Supplementary Figure 2; a methodology very similar to[17].

**Texel Array Experiment:** The experiment was carried out on a stripboard-based discrete component implementation of a 4-point texel array. The common load resistance was 300kΩ and the power supply

1.3V. Memristive devices were used as the memristor elements in each texel. All four devices used in the experiment resided on a single die with 32 available devices in total. Signals were fed into this system using two, dual-channel benchtop power supplies with two significant decimal digits resolution.

The benefit of using synthetic neural recording input data is that it contains ground truth information on spike identification and timing. On that basis an automatic sample selection script was ran on each spike instance available in the dataset in order to choose which data-samples from each instance are to be fed into the texel array for matching against a stored template. The script operated as follows: the data-points in each spike instance were read sequentially and once a trigger threshold $V_{trig}$ was exceeded for the first time the script skipped six samples and then choose the subsequent four as candidate inputs for the texel array set-up. This methodology was chosen because it rendered the three classes of spikes visibly distinguishable despite the use of only four template points. The overall 'trigger and sample' approach is similar to the work by Restituto-Delgado et al.[24]. In a more mature system implementation a larger texel array containing more than four samples would be used. Next, the extracted candidate four-texel sample sets were separated by single unit-template class. From each class, three texel sets were chosen for further processing: one featuring typical (M), one featuring lower than usual (L) and one featuring higher than usual (H) voltage values (selection shown in Supplementary Figure 8). Waveforms where the presence of more than one spike within each instance had corrupted the output of the sample selection script were automatically excluded from the selection. The voltage range of all nine selected sample sets (L, M, H instances for each of the three classes) was then adjusted by application of a common pair of gain and offset settings (Gain: 0.1; Offset: 0.66V). The adjusted texel data-point voltages were then suitable for working with the input voltage values the texel circuits were built to discriminate between. These adjusted values are shown in the inset of Figure 3(e) and were used as the input to the texel array after being rounded to 10mV precision (two significant decimal digits). This procedure caused the rounded texel voltages of the H instance of class 1 and the L instance of class 2 to completely overlap, hence that experiment was conducted only once for both cases.

**Data Availability:** The data that support the findings of this study are available from the corresponding author upon request, as detailed in http://www.nature.com/authors/policies/data/data-availablity-statements-data-citations.pdf.

**Supplementary Information** is available in the online version of the paper.

**Acknowledgements** We acknowledge the financial support of FP7 RAMP and EPSRC EP/K017829/1. The datasets used for the experiments in the manuscript are publicly available from the R. Q. Quiroga group at the university of Leicester (dataset 2).

**Author contributions** I.G., A.S. and T.P. conceived the experiments. A.K. and M.T. optimised and fabricated the devices. I.G. and A.S. carried out the experiments and performed the data analysis. All authors contributed towards writing the manuscript.

**Author Information** Reprints and permissions information is available at… The authors declare no competing financial interests. Correspondence and requests for materials should be addressed to I.G. (I.Gupta@soton.ac.uk)

# Fig. Legends

**Figure 1:** Memristive devices and neural signals. (a) Simplified schematic of experimental set-up. The top electrode of each test device was grounded whilst suitably amplified neural signal data samples were applied to the bottom electrode. Inset: Atomic Force Microscopy image memristive devices. (b, d, f) Neural signal data used as input to the memristive spike sorter. Each panel corresponds to neural spikes generated by different neurons and consequently featuring different signature waveforms. Thick traces: average spike waveform for each class (average of 10 instances). Thin traces: 10 different individual instances of spikes (c,e,g). Data includes amplification as shown in (a). Response of memristive devices to inputs from (b,d,f) respectively. Memristor resistive state jumps are observed in tight correlation to input signal voltage peaks. Pink dots: measured resistive state values. Blue: spike type/class I. Red: type II. Green: type III.

**Figure 2:** Spike sorting using memristive devices. (a) Input waveform for repeatability experiment. The waveform contains 10 identical copies of a spike triplet. Each triplet contains a succession of three different spike waveforms and is terminated by a reset pulse (example highlighted in yellow). Averaged spike waveforms were used in this case (thick traces in Figure 1(b,d,f)). Arrows and numbers indicate the timing and waveform class of each spike within a triplet. (b) Summary of results. Change in resistive state vs. starting resistive state for each pair of consecutive measurements taken whilst the input from (a) was applied to the memristive device (see Figure 1(c,e,g) and main text for details). The emergence of three distinct clusters of data-points corresponding to the three input spike classes is observed. Highlighted data-points (black outline): points gathered while applying the $8^{th}$ triplet as input. (c) Input waveform consisting of ten triplets, each constituted by random spike instances appearing in randomised order. (d) Corresponding results. The variability introduced by the input waveforms scatters the clusters which, however, remain broadly linearly separable. Boxed data point: the spike instance waveform contained a double spike (spike triplet 6 – highlighted in (c)).

**Figure 3:** Analogue domain template matching using memristive technologies. (a) Schematic of 'texel' circuit illustrating the breakdown into an enhanced inverter containing a memristor (R1) and the read-out stage. (b-c) Texel transfer characteristics from input voltage (b), through mid-point voltage $V_{MID}$ (c) and voltage at the output node (c). (e) Selected spike waveforms used as input to the test texel array. Crosses indicate the sample points used to feed the array. k: sample number. $V_{trig}$: texel array sampling trigger level (see methods section for description of sampling strategy). Inset: close-up of the chosen sample points. L, M, H and arrows: Low, medium and high voltage instances of spikes in class 3. (f) Schematic of 4-texel array used to carry out experiments. (g) Measured output voltage when spike samples from (e) are applied to the texel array in (f). Low, medium and high voltage (L, M, H) versions of spikes in each class shown. Higher voltage means greater degree of matching between input data and stored template. The texel array was programmed to respond best to class 2 spikes. Colours as in (e). Class 1-H and class 2-L results refer to the same experiment (discussed in methods section).

**Figure 1**

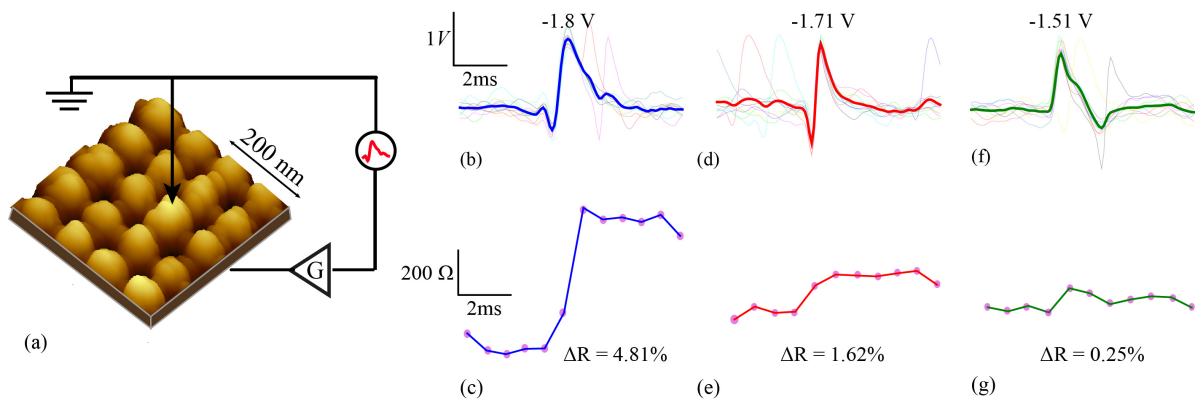

**Figure 2**

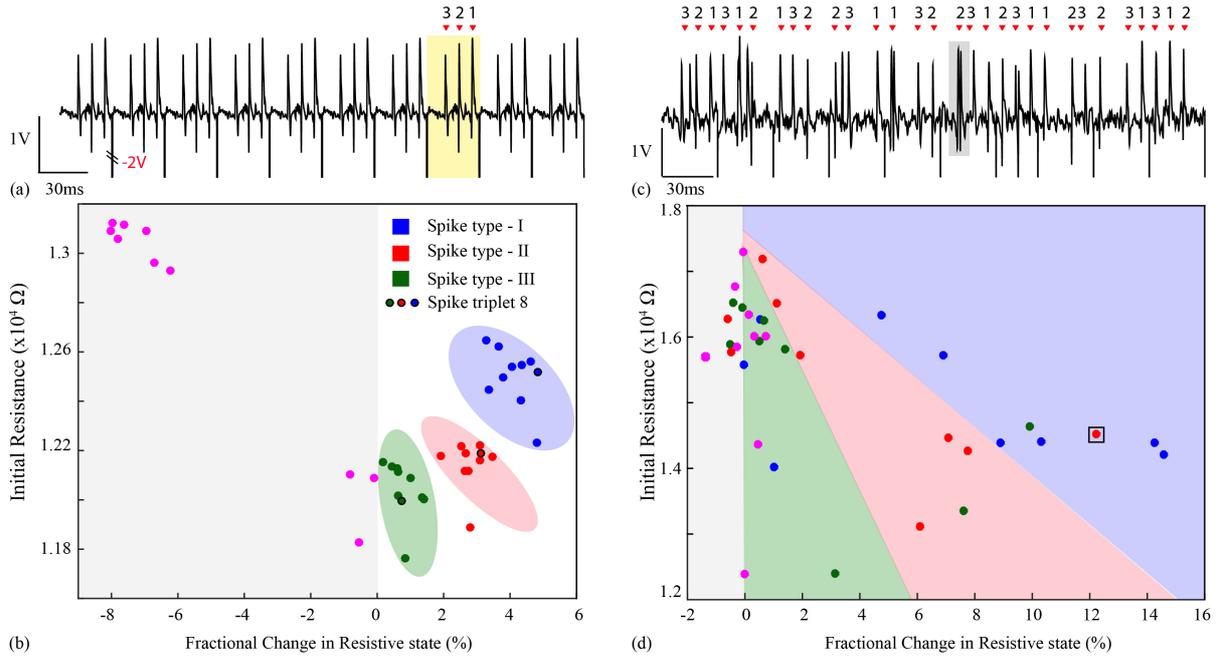

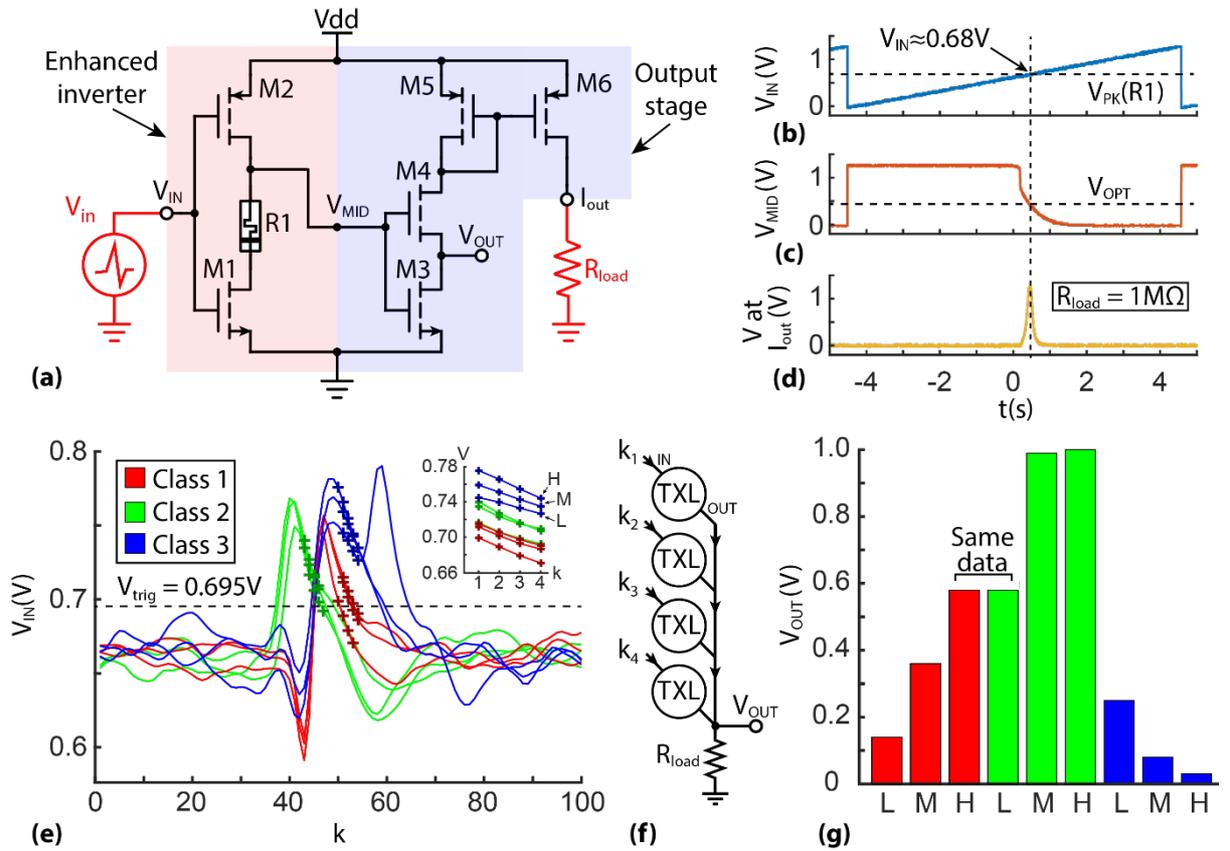

**Figure 3**

# Supplementary Information

# Spike sorting using non-volatile metal-oxide memristors

Isha Gupta[a,*], Alexantrou Serb[a], Ali Khiat[a], Maria Trapatseli[a], Themistoklis Prodromakis[a].

[a]Department of Electronics and Computer Science, Faculty of Physical Science and Engineering, University of Southampton, University Road, SO17 1BJ, Southampton, United Kingdom. Corresponding Author[*]: Isha Gupta (Email: I.Gupta@soton.ac.uk)

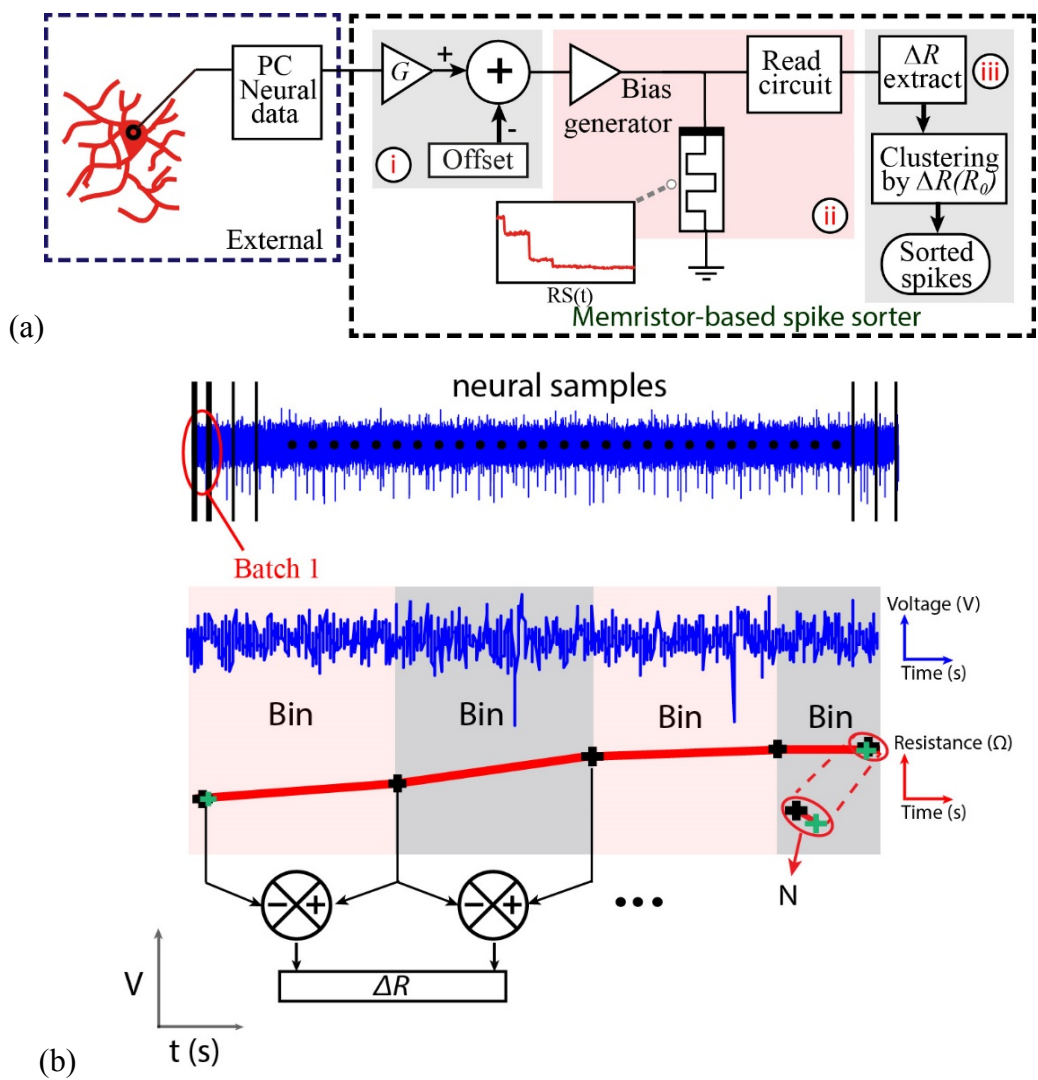

(a)

(b)

**Supplementary Figure 1**[1,2] A simplified system schematic for the memristor based spike-sorter[1]. (a) The setup for acquiring/recording neural data (for instance a CMOS based front-end system) is external to the presented memristor-based spike-sorter[1]. In this work, neural input data used for the experiments was stored on a PC. Data entering the memristor-based spike-sorter is first subjected to amplification and offset (i). The appropriate degree of amplification is determined following the thresholded integrator property of memristive

devices: significant neural activity (spikes) must lie above the switching threshold of the device whilst noise should remain below it[3]. In the second stage (ii), the conditioned waveforms are fed to the memristive devices using the characterisation instrument and the resistive state of the device is read periodically. The presence of supra-threshold spikes in the input is expected to cause significant changes in the resistive state of the device whilst the background noise will be inherently suppressed. Finally (iii), all captured resistive states are post-processed in order to estimate the number and identities of the spikes present.

(b) Signal-processing strategy used for the memristor-based spike sorter[1]. The neural signal is fed to the devices in batches. The batches are further divided in smaller bins and the resistive state of the device is read after each bin. Batch and bin sizes are flexible and can be set by the user to suit experimental needs[4]. Importantly, after each batch, the neural recording is paused and one additional resistive state measurement is taken in order to estimate noise levels (green cross in (b)). Therefore, during post-processing, consecutive resistive state readings are used to estimate the changes in resistive state of the device and the readings at the end of each batch are used to estimate the noise. For spike sorting, the changes in resistive state registered in each bin are plotted as a function of the initial resistive state for each bin.

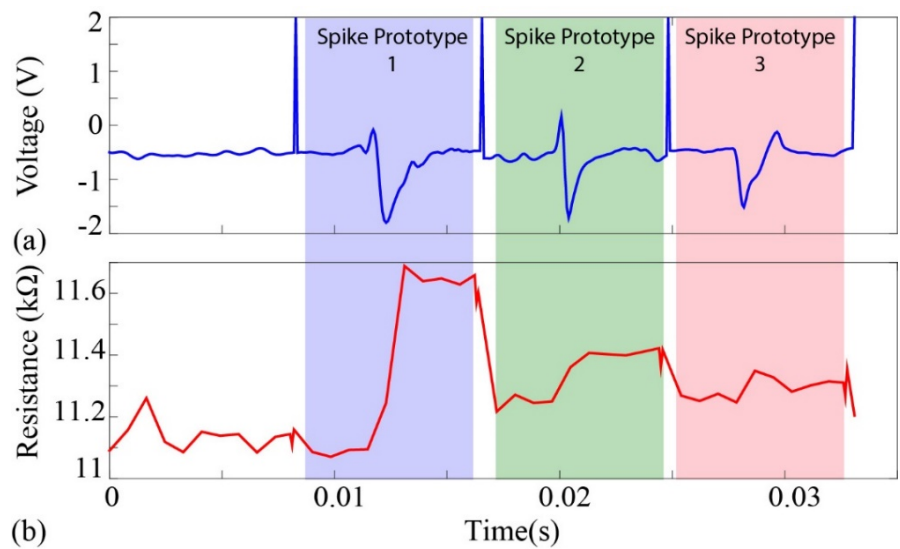

**Supplementary Figure 2** (a) Pre-processed recording used for Figure 1(b-g) in the main manuscript. The gain and amplification used for the recording was -1.31 and . Reset pulses of +2V were used to maintain the device functional within the operational resistive state region. (b) Resistive state changes for the device in response to the recording in (a) over time. Blue, green and red shadings correspond to the three distinct average prototypes 1, 2 and 3 respectively.

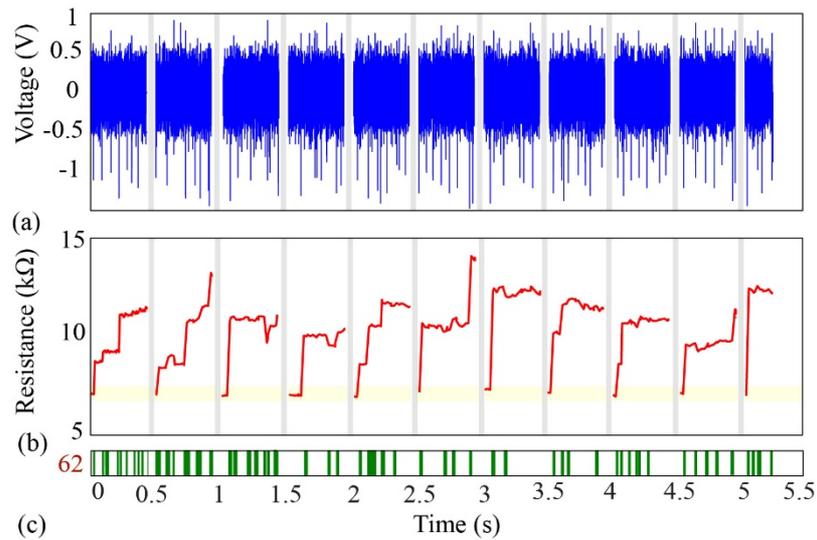

**Supplementary Figure 3**[2] Frequent resetting of the memristive device (non-volatile regime[2]). Continuous operation of the memristive device in the non-volatile region for input signals dominated by peaks of a single polarity results in saturation of the resistive state of the device. As a mitigation measure, reset pulses shown in grey band are interspersed with the neural signal. (a), (b) Resistive state response of the device-under-test in response to the sub-neural recordings. After every sub-neural recording the device is reset to its intital resistive state (yellow bands) using a pulse of positive polarity of 100 µs. The intital resistive state of the device is in the region of 6-8 kΩ and the operation of the device is in the region of 6 kΩ (low resistive state) to 15 kΩ (high resistive state). (c) Raster of detected spikes. Total spike count: 62.

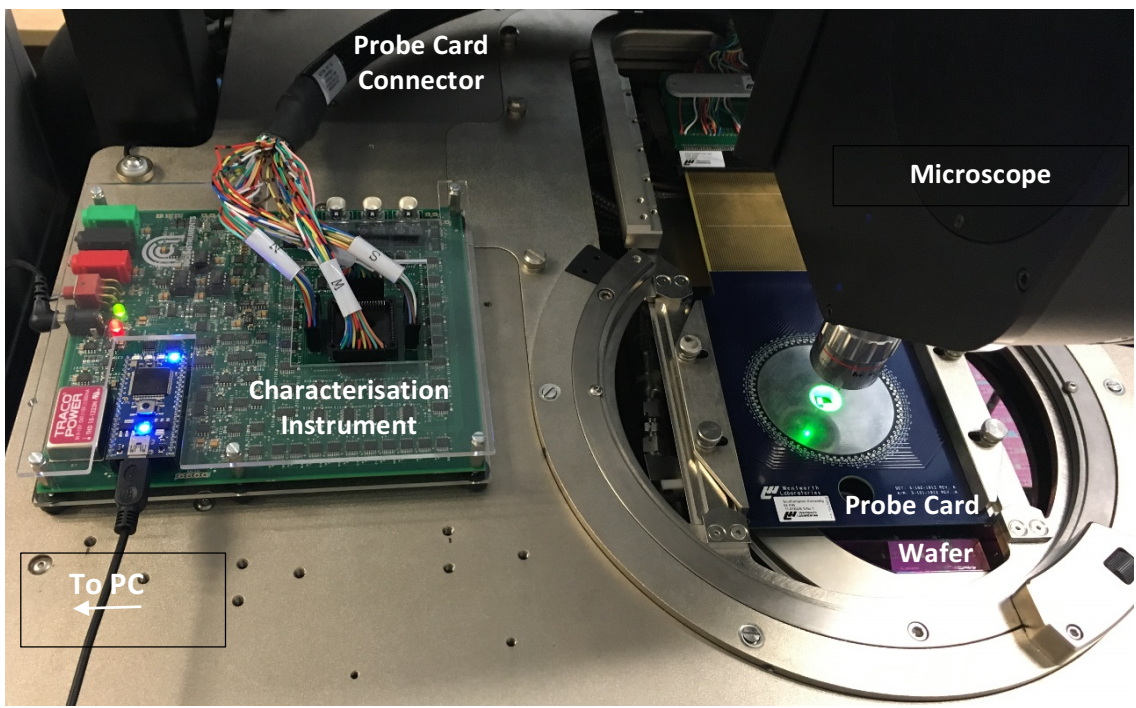

**Supplementary Figure 4** Experimental setup for the spike sorting experiments. The devices were electrically characterised using a custom-made hardware characterisation instrument[5].

The instrument can be used to characterise devices both in-package and directly on-wafer via probe card. For the demonstrated experiments, both packaged and on-wafer devices were used[6,1].

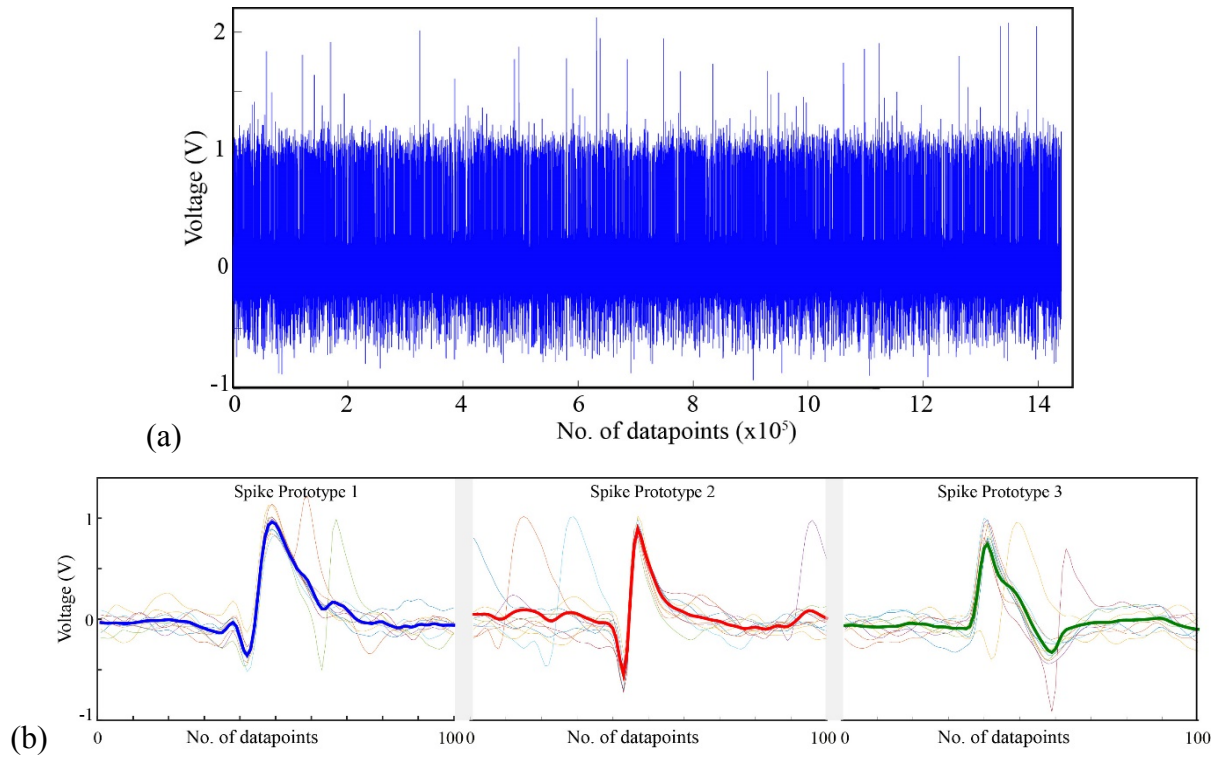

**Supplementary Figure 5** (a) Raw neural recording data used for the experiments. The data is publicly available from University of Leicester (http://www2.le.ac.uk/centres/csn/software, Dataset 2[7]. The neural recording contains three distinct single-unit spike waveforms superposed on background noise. Detailed description on how this neural recording data was synthesised is presented in the reference[7]. (b) For our experiments, we randomly extracted ten different instances of each spike prototype as illustrated in the three insets. The spike timings of the instances are documented in Supplementary Table 1. Each instance of every spike waveform contains 100 data points i.e. 19 points before and 80 points after the spike registration timestamp. Thick blue, green and red traces show the averages of ten different instances of each spike waveform. The maximum and minimum voltage values in each average spike prototype are further presented in Supplementary Table 2.

| S.No. | Spike Waveform 1 (Spike timings) | Spike Waveform 2 (Spike timings) | Spike Waveform 3 (Spike timings) |
|---|---|---|---|
| 1. | 2785 | 20173 | 45450 |
| 2. | 54871 | 41370 | 465326 |
| 3. | 127642 | 107486 | 606816 |
| 4. | 207821 | 194634 | 854280 |
| 5. | 395160 | 300366 | 939685 |
| 6. | 708914 | 395161 | 1091743 |
| 7. | 903698 | 631599 | 1208267 |
| 8. | 1011435 | 886880 | 1314658 |
| 9. | 1273906 | 1082916 | 1436188 |
| 10. | 1420564 | 1330703 | 204139 |

**Supplementary Table 1** Spike timings for ten randomly chosen instances of three distinct spike prototypes (as shown in Supplementary Figure 5(b)). Units: data point indices.

| S.No. | Average Spike Prototypes | Max. (V) | Min. (V) |
|---|---|---|---|
| 1. | Spike Prototype 1 | 1.042 | -0.33 |
| 2. | Spike Prototype 2 | 0.968 | -0.52 |
| 3. | Spike Prototype 3 | 0.850 | -0.30 |

**Supplementary Table 2** Maximum and minimum voltage values in the three average spike prototype waveforms as illustrated in Supplementary Figure 5(b).

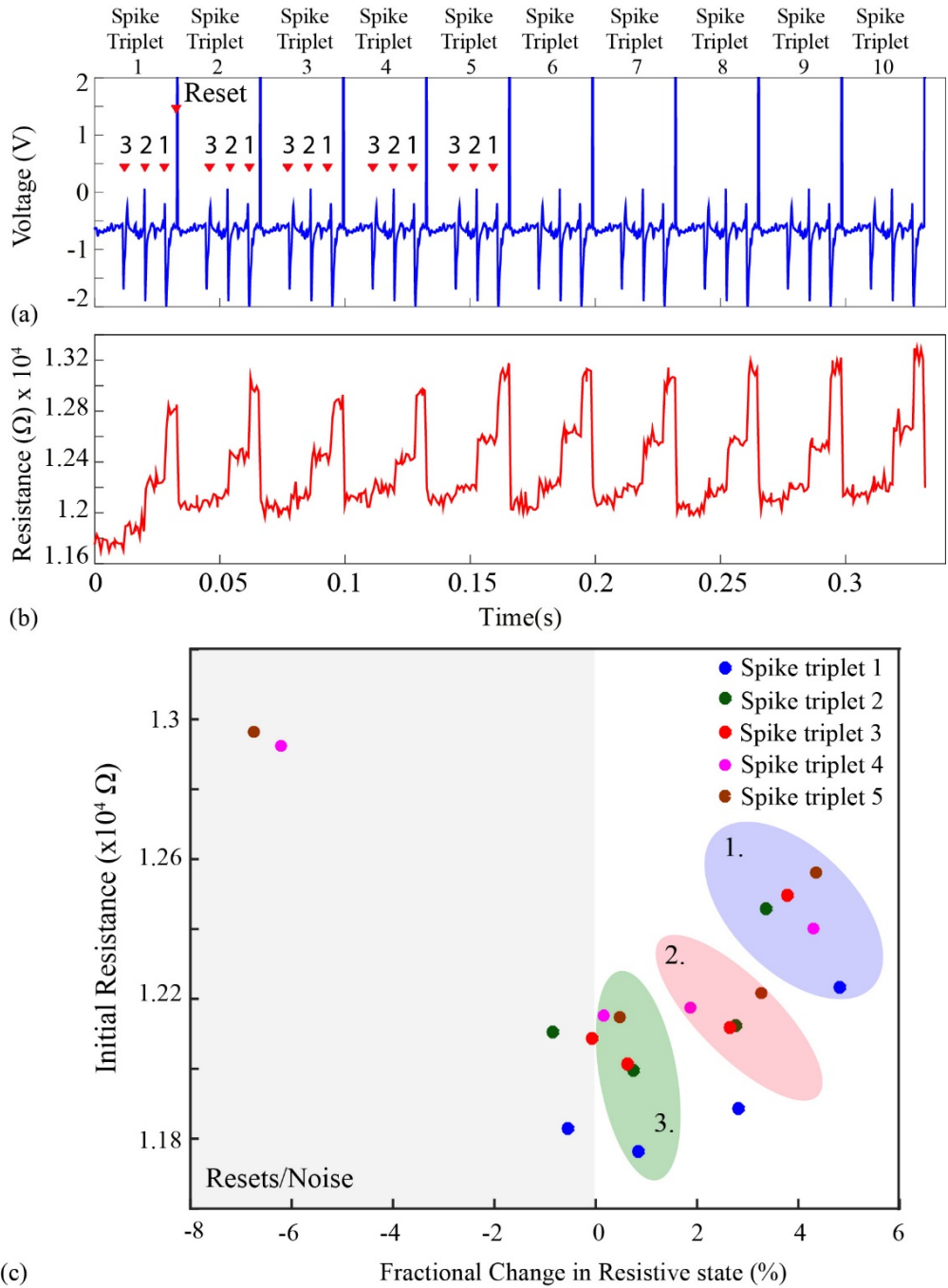

**Supplementary Figure 6** (a) Input signal used for Figure 2(a) in the main manuscript. Ten copies of a spike triplet were concatenated and fed to the device-under-test. The triplet was composed of the three average spike prototypes shown in Supplementary Figure 5(b), arranged in increasing order of amplitude i.e. order 3, 2, 1. Reset pulses followed every spike triplet. This stimulus was chosen in order to experimentally demonstrate the response repeatability of the device-under-test. (b) Resistive state evolution of the device in response to the neural recording in (a). (c) Resistive state change between every pair of consecutive measurements plotted as a function of the first resistive state measurement for the same pair (please see Supplementary Figure 1 and Supplementary Table 3 for further explanation). Data points are colour-coded for the corresponding triplet.

| S.No. | Resistive State Reads (RSR's - Ω) | Batch 1 | RSR's | Batch 2 | RSR's | Batch 3 | RSR's | Batch 4 |
|---|---|---|---|---|---|---|---|---|
| | Background/Resets | | Spike III | | Spike II | | Spike I | |
| 1. | 11757.99316 | B1 | 11782.02 | B13 | 11898.06 | B25 | 12311.09 | B37 |
| 2. | 11828.51953 | B2 | 11763.65 | B14 | 11886.4 | B26 | 12232.41 | B38 |
| 3. | 11790.73242 | B3 | 11757.92 | B15 | 11809.14 | B27 | 12235.02 | B39 |
| 4. | 11699.33887 | B4 | 11760.06 | B16 | 11983.17 | B28 | 12241.67 | B40 |
| 5. | 11790.91992 | B5 | 11706.08 | B17 | 11855.17 | B29 | 12267.59 | B41 |
| 6. | 11790.03516 | B6 | 11894.95 | B18 | 12191.47 | B30 | 12653.33 | B42 |
| 7. | 11783.03027 | B7 | 11899.9 | B19 | 12244.21 | B31 | 12843.73 | B43 |
| 8. | 11697.71387 | B8 | 11924.52 | B20 | 12283.07 | B32 | 12776.11 | B44 |
| 9. | 11755.08691 | B9 | 11834.5 | B21 | 12238.37 | B33 | 12796.08 | B45 |
| 10. | 11766.31055 | B10 | 11841.8 | B22 | 12179.92 | B34 | 12816.69 | B46 |
| 11. | 11739.83008 | B11 | 11889.99 | B23 | 12279.27 | B35 | 12813.46 | B47 |
| 12. | 11784.81836 | B12 | 11855.77 | B24 | 12202.14 | B36 | 12834.55 | B48 |
| Fractional Changes in Resistive state – ignoring data point at S. No. = 1 | | | | | | | | |
| S.No. (10-2) | -0.52592367 | | 0.664288 | | 2.469372 | | 4.776493 | |
| S.No. (11-2) | -0.749793351 | | 1.073951 | | 3.305206 | | 4.7501 | |
| S.No. (12-2) | -0.369455974 | | 0.783066 | | 2.656249 | | 4.922454 | |
| Averages | -0.548390998 | | 0.840435 | | 2.810276 | | 4.816349 | |
| Fractional Change in Resistive state – including data point at S. No. = 1 | | | | | | | | |
| S.No. (10-1) | 0.070738 | | 0.507369 | | 2.368994 | | 4.106838 | |
| S.No. (11-1) | -0.15447 | | 0.916394 | | 3.204009 | | 4.080613 | |
| S.No. (12-1) | 0.228144 | | 0.625962 | | 2.55687 | | 4.251865 | |
| Average | 0.048136 | | 0.683242 | | 2.709563 | | 4.146439 | |

**Supplementary Table 3** Raw resistive state measurements taken during the first spike triplet shown in Supplementary Figure 6(a,b). Each resistive state read-out is indexed with a unique identifier ('B1, B2, …'). Each batch contains 100 input data points and each bin contains 10 input data points. This means that resistive state measurements were taken at the beginning of each batch of size 100 and then after every bin of size 10. One final measurement was taken at the end of batch whilst the neural feed was paused (noise estimation – see Supplementary Figure 1). **Important:** (a) For estimating the fractional change in resistive states for each spike prototype in the spike triplet, resistive states highlighted in yellow and pink were used as the initial and final resistive states respectively. (b) For estimating fractional changes in resistive states due to background noise/resets, resistive states indicated in yellow and peach colours (S.No. 11/12) were used as the initial and final resistive states respectively. The averages of these values were plotted in Figure 2(b,d) and Supplementary Figure 6(c) as a function of the initial resistive state.

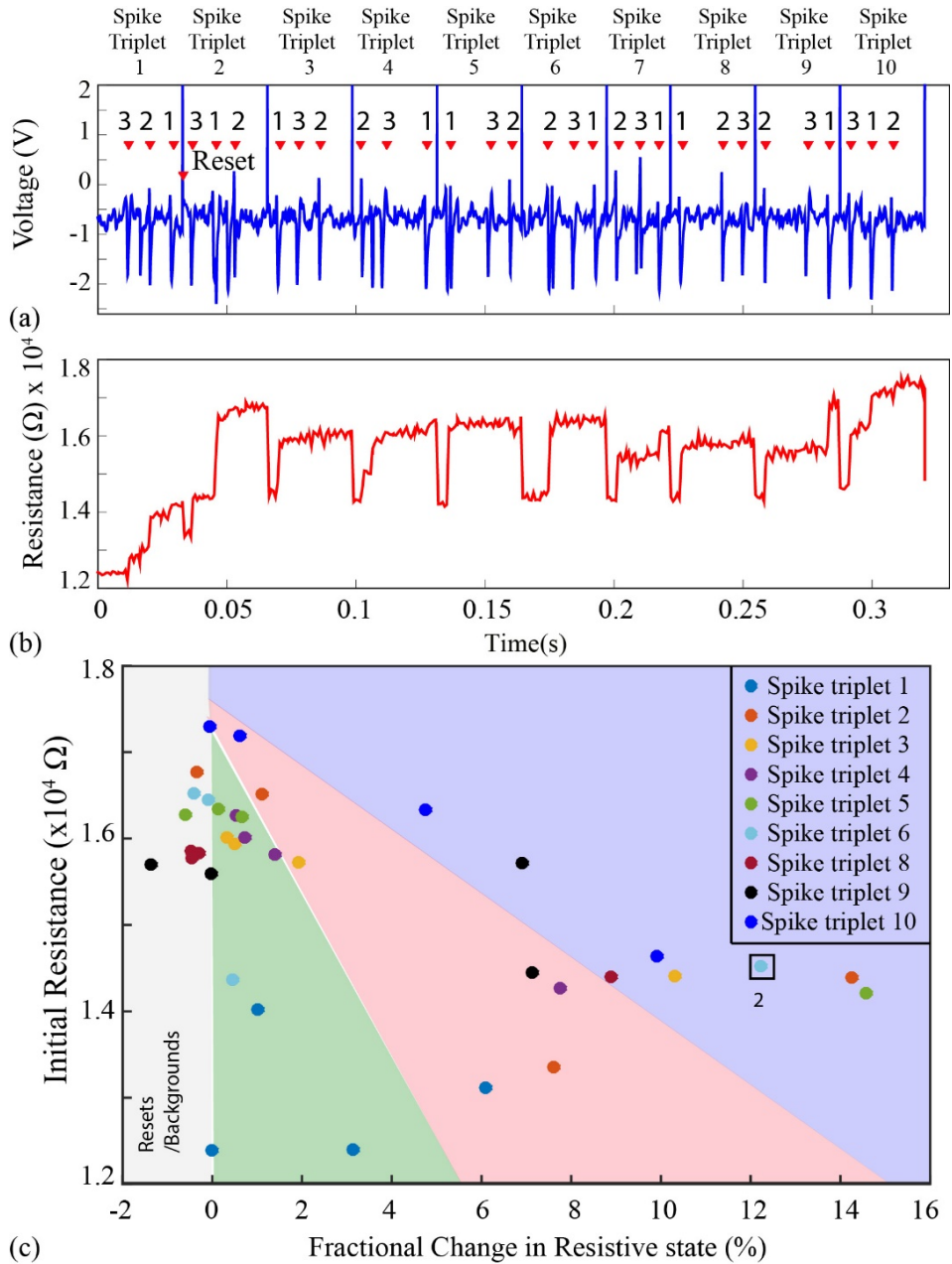

**Supplementary Figure 7** (a) Neural waveform used for Figure 2(c) in the main manuscript. Randomly chosen spike waveform instances, one belonging to each prototype, were arranged in random order as shown in the figure in order to form a spike triplet. Reset pulses followed every spike triplet. Ten such randomly generated triplets were concatenated and fed to the device-under-test. (b) Resistive state evolution of the device in response to the neural recording in (a). (c) Resistive state change between every pair of consecutive measurements plotted against the first resistive state measurement for each corresponding pair (please see Supplementary Figure 1(b) and Supplementary Table 4 for further explanation.).

|  | \multicolumn{2}{c}{Background Resets} | \multicolumn{2}{c}{Spike III} | \multicolumn{2}{c}{Spike II} | \multicolumn{2}{c}{Spike I} |
|---|---|---|---|---|---|---|---|---|
| Spike Triplet (Order fed) | RSR (% change) | Initial RS (Ω) | RSR (% change) | Initial RS (Ω) | RSR (% change) | Initial RS (Ω) | RSR (% change) | Initial RS (Ω) |
| Spike Triplet 1 (3,2,1) | -0.00895 | 12392.02 | 3.131351 | 12400.28 | 6.083045 | 13111.9 | 1.012399 | 14022.04 |
| Spike Triplet 2 (3,1,2) | -0.34793 | 16769.94 | 7.612269 | 13351.12 | 1.10533 | 16514.83 | 14.24979 | 14387.25 |
| Spike Triplet 3 (1,3,2) | 0.331 | 16012.95 | 0.492714 | 15939.68 | 1.922429 | 15721.62 | 10.30495 | 14407.54 |
| Spike Triplet 4 (2,3,1) | 0.523822 | 16265.24 | 1.40102 | 15812.35 | 7.751934 | 14267.21 | 0.719231 | 16012.79 |
| Spike Triplet 5 (1,3,2) | -0.60705 | 16277.72 | 0.134587 | 16337.895 | 0.663175 | 16246.95 | 14.57128 | 0.45586 |
| Spike Triplet 6 (2,3,1) | -0.09134 | 16451.23 | -0.41374 | 16518.9 | 0.45586 | 14365.22 | 12.2216 | 14525.54 |
| Spike Triplet 8 (1,2,3) | -0.47188 | 15875.08 | -0.45154 | 15775.74 | -0.29803 | 15827.51 | 8.88117 | 14395.29 |
| Spike Triplet 9 (2,3,1) | -0.02517 | 15591.67 | -1.37581 | 15698.12 | 7.125367 | 14452.56 | 6.900 | 15719.95 |
| Spike Triplet 10 (3,1,2) | -0.05287 | 17298.12 | 9.900594 | 14640.5 | 0.61479 | 17190.34 | 4.75116 | 16335.51 |

**Supplementary Table 4:** Result summary for Supplementary Figure 7. Measured resistive states and fractional changes in resistive state experienced by the memristor as a result of applying the ten spike triplets from Supplementary Figure 7(a). In each row, the effects of the different spike components of each triplet are shown separately.

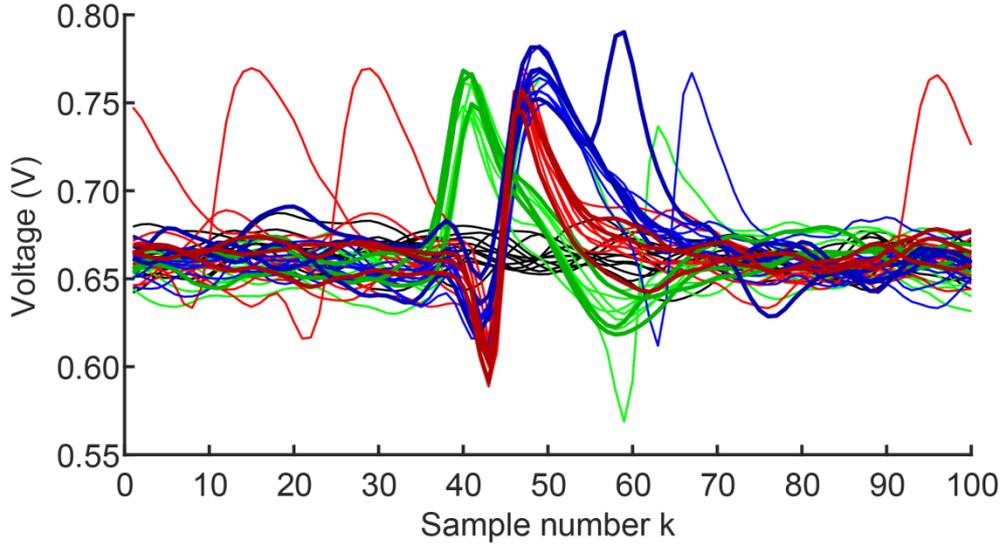

**Supplementary figure 8:** Dataset used for carrying out texel array experiment in Figure 3. Shown are all neural spike waveforms included in the dataset with colours indicating their class (same colour scheme as Figure 3). Spike waveforms chosen as inputs for the experiment in Figure 3 are shown as thicker, darker traces.

|  | SPIKE CLASS | TXL1 | TXL2 | TXL3 | TXL4 | OUTPUT (1st run) | OUTPUT (2nd RUN) |
|---|---|---|---|---|---|---|---|
| **Ideal** | 3H | 0.7757 | 0.7657 | 0.7546 | 0.7443 | 0.03 | 0.06 |
| **Rounded** |  | 0.78 | 0.77 | 0.76 | 0.74 |  |  |
| **Ideal** | 3M | 0.7592 | 0.7510 | 0.7427 | 0.7350 | 0.08 | 0.16 |
| **Rounded** |  | 0.76 | 0.75 | 0.74 | 0.74 |  |  |
| **Ideal** | 3L | 0.7450 | 0.7397 | 0.7329 | 0.7266 | 0.25 | 0.39 |
| **Rounded** |  | 0.75 | 0.74 | 0.73 | 0.73 |  |  |
| **Ideal** | 2H | 0.7400 | 0.7270 | 0.7163 | 0.7071 | 1.00 | 1.19 |
| **Rounded** |  | 0.74 | 0.73 | 0.72 | 0.71 |  |  |
| **Ideal** | 2M | 0.7347 | 0.7231 | 0.7149 | 0.7094 | 0.99 | 1.23 |
| **Rounded** |  | 0.73 | 0.72 | 0.71 | 0.71 |  |  |
| **Ideal** | 2L | 0.7164 | 0.7058 | 0.6983 | 0.6923 | 0.58 | 0.83 |
| **Rounded** |  | 0.72 | 0.71 | 0.70 | 0.69 |  |  |
| **Ideal** | 1H | 0.7151 | 0.7055 | 0.6971 | 0.6904 |  |  |
| **Rounded** |  | 0.72 | 0.71 | 0.70 | 0.69 |  |  |
| **Ideal** | 1M | 0.7115 | 0.7014 | 0.6926 | 0.6863 | 0.36 | 0.62 |
| **Rounded** |  | 0.71 | 0.70 | 0.69 | 0.69 |  |  |
| **Ideal** | 1L | 0.6991 | 0.6888 | 0.6788 | 0.6705 | 0.14 | 0.28 |
| **Rounded** |  | 0.70 | 0.69 | 0.68 | 0.67 |  |  |

**Supplementary table 5:** Results for texel array experiment. Ideal (computed) and rounded values used as voltage inputs to the texel array elements (TXL1-4) are shown alongside the resulting output voltages at node $V_{OUT}$ for two repetitions of the experiment. All units are Volts.

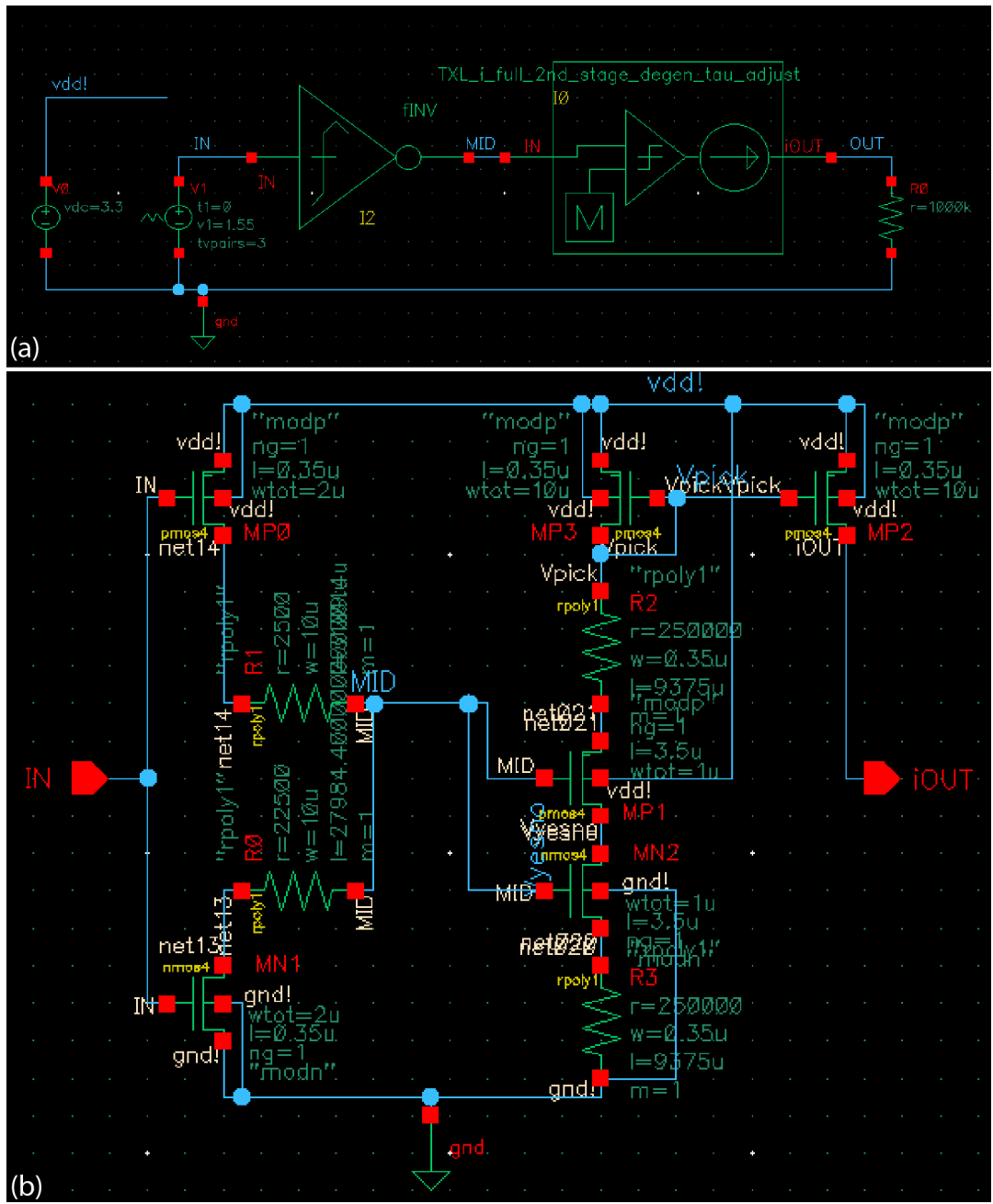

**Supplementary figure 9:** Schematics of texel power dissipation test bench (a) and texel circuitry (b) used to carry out power dissipation simulations. The driving inverter in (a) is similar to the inverter in (b), i.e. devices MP0, MN1, R0 and R1. In (b) memristors are represented by resistive elements. R0 and R1 are the memristors tuning the transfer characteristics of the texel whilst R2 and R3 are optionally implemented for tuning the sensitivity of the output current to input voltage and to act as current limiters. These memristors need not switch after fabrication and act as simple resistive loads.

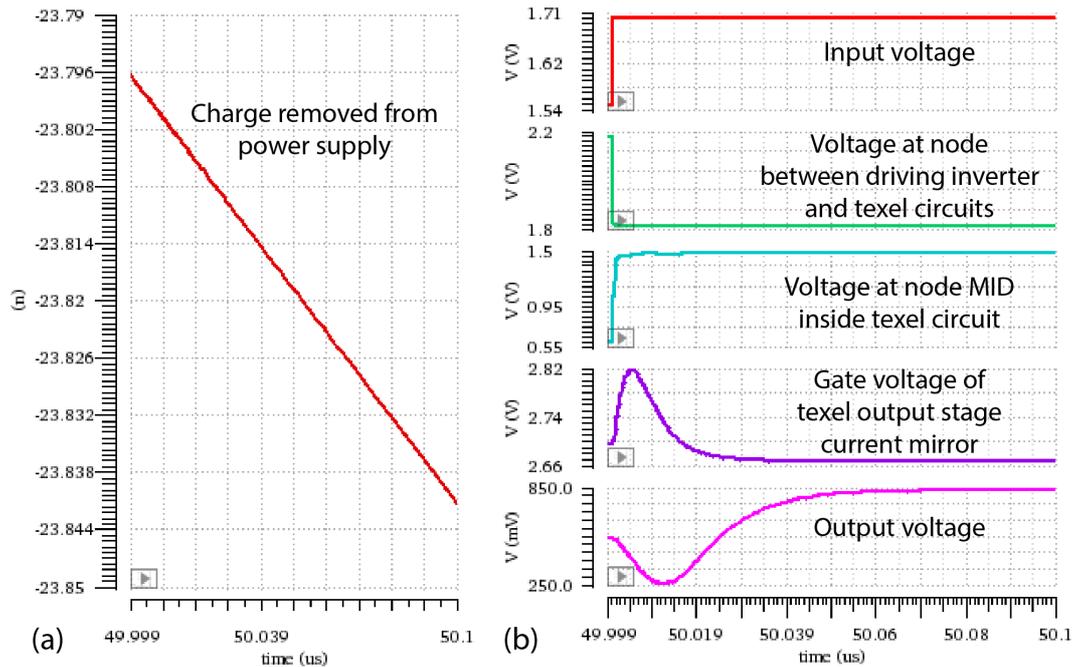

**Supplementary figure 10:** Charge dissipation of the test bench circuit in Supplementary Figure 9 for input signal transition from 1.55V to 1.70V. AMS 0.35 micron technology with power supply set to 3.3V. Approximately 46fC escape the power supply throughout the process, corresponding to toggling 37 minimum drive strength inverters as shown in Supplementary Figure 10. The design under study has not been optimised for power. (a) Charge dissipation through test circuit over time. (b) Selected voltage signal time evolutions from system input (red trace) to system output (green trace).

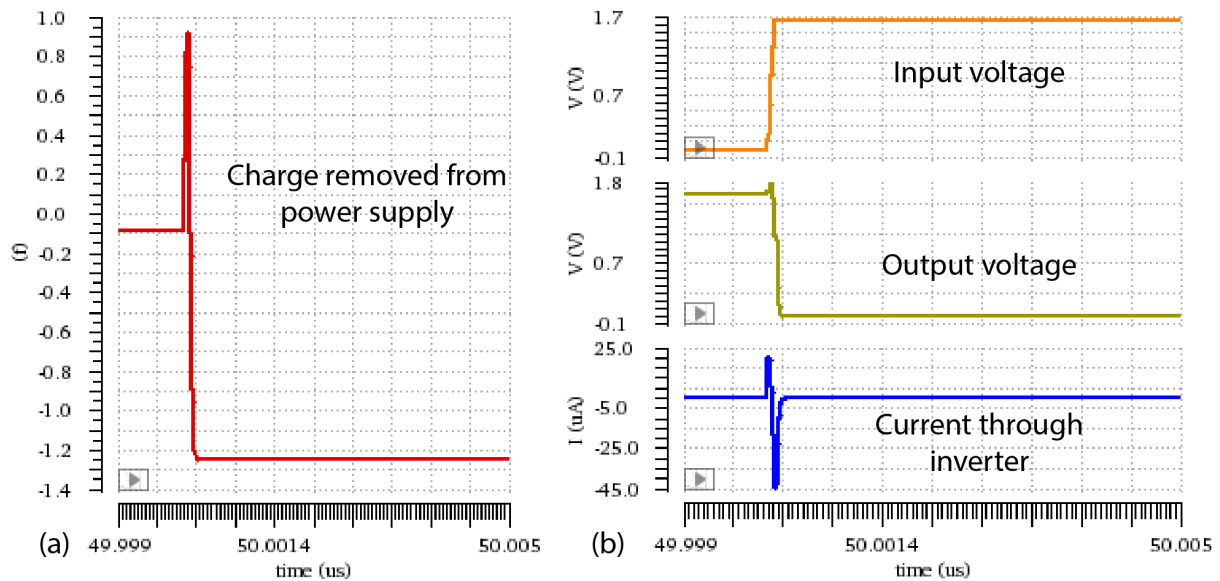

**Supplementary figure 11:** Charge dissipation of a minimum strength inverter in AMS 0.35 micron technology for a single input signal toggle. Approximately 1.25fC escape the power supply through the process. VDD = 1.65V. (a) Charge removed from power supply vs. time. (b) Input voltage signal. (c) Output voltage. (d) Current through the inverter.

|  | TXL1 | TXL2 | TXL3 | TXL4 |
|---|---|---|---|---|
| **Before run** | 19.5k | 14.8k | 12.7k | 10.6k |
| **After run** | 19.5k | 15.1k | 12.6k | 10.6k |

**Supplementary table 6:** Resistive states of memristors used for texel array experiment, second round.

**Supplementary material 1: Power estimations for texel circuit operation.**

In order to estimate the power budget under which a texel can operate, industry-standard Cadence tool-based simulations were ran. A single texel was simulated (schematic in Supplementary Figure 9(b)) within the power dissipation estimation testbench shown in Supplementary Figure 9(a). The reference technology was AMS 0.35 micron. Resistors were used to model memristors, and the amount of charge removed from the power supply to carry out the computation was taken as a proxy for power dissipation.

Operating power estimations were benchmarked for the following analogue computation: input voltage rises from 1.55V to 1.7V. These values guaranteed a visible change in the system output voltage level, as evidenced in Supplementary Figure 10(b). The simulator indicates that by the time the output voltage stabilises the overall amount of charge removed from the power supply is approximately 46fC. This compares favourably with the ~1.25fC charge dissipated by an industrially-designed minimum size inverter for a single digital state transition (input 0 to 1) in the same technology, as shown in Supplementary Figure 11. Therefore for the charge dissipation price of ~39 inverter toggles the texel carries out an analogue input-output mapping operation. Notably, the texel circuit used in this simulation was not optimised for low power dissipation but is provided as a working example that can be set up with minimum design effort.

**Supplementary video 1**: All four texels in the array work normally. The video begins by showing the texel array stripboard (bottom left) receiving inputs from the dual channel power supplies (0.70V, 0.69V, 0.68V and 0.67V for each channel respectively as seen on the green 7-segment displays of the power supplies). The memristive chip can be seen sitting on its own PCB right next to the strip-board, connecting to the strip-board via red and black wires. The oscilloscope's purples trace shows the voltage at the output node ($V_{OUT}$ in Fig. 3(f)). The starting input voltages are all below the preferred input voltages of the four texels. As the video progresses the voltage on each channel is manually increased until it lies firmly above the preferred input voltage for each texel. The output voltage shows an increase followed by a decrease each time this procedure is carried out, thus proving that each texel's preferred input voltage lies between the initial and final values of the voltages displayed on the 7-segment displays of the power supplies.